\documentclass[sigconf, nonacm]{acmart}

\usepackage{algorithm}
\usepackage{algpseudocode}
\AtBeginDocument{%
  }

\begin{document}

\title{Whose Narrative is it Anyway?  A KV Cache Manipulation Attack}

%

\author{Mukkesh Ganesh}
\affiliation{%
  \country{USA}
  }
\email{mganesh@usc.edu}

\author{Kaushik Iyer}
\affiliation{%
  \country{USA}
  }
\email{kaushikn@usc.edu}

\author{Arun Baalaaji Sankar Ananthan}
\affiliation{%
  \country{USA}
  }

\email{arunbaalaajis@alumni.usc.edu}

\begin{abstract}
The Key Value(KV) cache is an important component for efficient inference in autoregressive Large Language Models (LLMs), but its role as a representation of the model's internal state makes it a potential target for integrity attacks. This paper introduces "History Swapping," a novel block-level attack that manipulates the KV cache to steer model generation without altering the user-facing prompt. The attack involves overwriting a contiguous segment of the active generation's cache with a precomputed cache from a different topic. We empirically evaluate this method across 324 configurations on the Qwen 3 family of models, analyzing the impact of timing, magnitude, and layer depth of the cache overwrite. Our findings reveal that only full-layer overwrites can successfully hijack the conversation's topic, leading to three distinct behaviors: immediate and persistent topic shift, partial recovery, or a delayed hijack. Furthermore, we observe that high-level structural plans are encoded early in the generation process and local discourse structure is maintained by the final layers of the model. This work demonstrates that the KV cache is a significant vector for security analysis, as it encodes not just context but also topic trajectory and structural planning, making it a powerful interface for manipulating model behavior.
\end{abstract}
\maketitle

\section{Introduction}
The emergence of Large Language Models (LLMs) as a transformative force for both the general public and major corporations hinges on their ability to provide low-latency inference. For autoregressive LLMs, this is achieved by relying on the key value cache(KVCache)\cite{KVCachePap} optimization to avoid the significant computational overhead of recomputing attention across the full context for every generated token. This optimization turns the expensive quadratic attention computation into a simple linear operation.
This has consequently made the cache a central component of modern serving systems such as vLLM\cite{Kwon:2023:PagedAttention}, Sarathi Serve\cite{Agrawal:2024:SarathiServe}, and LMCache\cite{Cheng:2025:LMCache}, which reuse and share cached prefixes to improve throughput.

At the same time, the cache has become a meaningful representation of a model's internal state during generation. This is because the K and V vectors it stores are not arbitrary but are the model's cumulative understanding of the entire preceding context, effectively serving as the "memory" that informs each subsequent token's generation.

Recent work on representation engineering\cite{wu2024reft} and activation steering has established that internal activations\cite{xu2025easysteer} are a practical interface for influencing model outputs , as modifying them can alter a model's behavior in controlled ways. Consequently, security research has predominantly concentrated on two types of threats: confidentiality attacks, which extract sensitive information from shared caches\cite{wu2025iknow}, and availability attacks, which introduce noise to impair output quality\cite{hossain2025can}. While these studies demonstrate that attackers can access cache data, they do not address the deliberate manipulation of the cache to influence or steer model generation.

This gap motivates our work. In this paper, we study the integrity of the KVCache and examine how models behave when an adversary overwrites part of their cached history with structured, unrelated data.
We propose History Swapping, a block-level cache modification attack in which a contiguous segment of the active generation history is replaced with precomputed values from a different topic. The model then continues text generation based on the modified cache, allowing the conversation to be steered without altering the visible prompt.

We evaluate this attack on the Qwen 3\cite{Qwen3:2025} model family and analyze how timing, magnitude, and layer location affect vulnerability.
Our findings indicate that overwrites in later layers can consistently redirect generation, whereas early-layer and partial-layer interventions are more robust to manipulation.
These findings highlight the cache as an important target for both security analysis and interpretability, since it encodes plan structure and topic direction that can persist even after a partial corruption.

\subsection{Our Contributions}
This paper presents three key contributions:
\begin{enumerate}
    \item We define History Swapping, a block-level integrity attack on the KV cache, and show that it can potentially steer ongoing conversations toward attacker-chosen topics across a range of settings.
    \item We introduce a parameterized framework for cache integrity analysis, varying timing (swap\_token), \\ magnitude (swap\_percent), and architectural depth/location (layers\_affect\_percent, from\_beginning).
    \item We empirically evaluated the Qwen 3 family, characterizing resilience and failure modes under History Swapping.
\end{enumerate}
\section{Related Work}
Work on the KV cache spans across systems efficiency, internal state manipulation, and security analysis.
Prior work has primarily considered the cache either as a performance optimization or a privacy vulnerability, but has not explored how deliberate, structured modifications can be used to steer model generation.

\subsection{Systems Optimization for KV Caches}
Modern LLM inference systems are built using KV cache to avoid recomputing attention over the full context.
Systems such as vLLM\cite{Kwon:2023:PagedAttention} and Sarathi Serve\cite{Agrawal:2024:SarathiServe} organize cached values into contiguous memory regions that can be allocated, paged, or reused across requests. LMCache\cite{Cheng:2025:LMCache} extends this approach by storing and sharing cached prefixes across sessions. These systems treat the cache as a sequence of discrete regions that align with consecutive tokens.
In this work a block level operation means replacing an entire contiguous span of cached time steps across layers, rather than modifying individual token vectors or adding perturbations within a single position.

\subsection{Manipulation of Internal States}
Prior work has shown that modifying intermediate representations can systematically influence model outputs. Techniques such as representation engineering manipulate latent spaces during fine-tuning to steer model behavior \cite{wu2024reft}, while activation steering introduces carefully designed activation vectors at inference to guide predictions without additional training \cite{xu2025easysteer} \cite{kirtania2025steering}. Collectively, these findings indicate that transformer models can tolerate significant alterations to hidden states while still producing coherent outputs.
Building on this insight, we propose History Swapping, a method that operates directly on the KV cache. Rather than applying small perturbations or retraining the model, History Swapping replaces an entire contiguous block of the cache with a coherent history segment derived from a different topic.

\subsection{Security Analyses of the KV Cache}
Security work on the KV cache has been primarily on two attack classes.
Confidentiality attacks, which infer sensitive prompts via cache timing \cite{wu2025iknow}\cite{chu2025selective}, and Availability attacks, which corrupt cached values to degrade outputs or induce repetition\cite{hossain2025can}. While these studies show that cache data is vulnerable, they are either observational or disruptive and do not examine the effects of intentionally inserting structured alternative histories.

\subsection{Gap and Positioning}
Where previous state manipulation methods operate through fine-tuning or injecting small activation offsets, History Swapping introduces a fundamentally different mechanism: the block-level replacement of a model's internal history.
This technique uses semantically coherent content to directly engage the model's memory, offering a powerful tool to probe its reliance on cached information for topic control, planning, and logical continuation.

\section{The History Swapping Methodology}
\subsection{Preliminaries}
Autoregressive LLMs compute each new token by attending to all previously generated key (K) and value (V) vectors.  
During generation, these vectors are appended to the KV cache so they do not need to be recomputed.  
At generation step $t$, the cache stores $K$ and $V$ for tokens $1$ through $t-1$, and the model appends $(k_t, v_t)$ after producing token $t$.

Because the cache encodes the model's evolving internal representation of the prompt and prior outputs, modifying it enables an attacker to influence future generations without altering the visible text.  
We define a \emph{block} as a contiguous set of time steps in the cache, consisting of the K and V vectors for consecutive tokens across all layers.

\subsection{Threat Model}
The attacker seeks to steer an active generation toward an attacker-chosen topic.  
The attacker cannot modify model weights or prompts but can read and write KV cache tensors either in storage or during inter-hardware transfers.

This capability arises in two deployment scenarios:
\begin{enumerate}
    \item \textbf{Stored cache access:} Systems that persist cached prefixes for reuse or sharing may allow an attacker to modify stored cache segments before the next decoding step.
    \item \textbf{In-flight access:} Hybrid TEE–accelerator setups transmit KV tensors between trusted and untrusted components. An attacker who intercepts these transfers can alter the tensors prior to use.
\end{enumerate}

The attacker’s goal is not disruption or extraction, but controlled insertion of an alternate internal history that shifts the model’s behavior.

\subsection{History Swapping Attack}
The core mechanism of the attack is shown in Algorithm~\ref{alg:history-swapping}.  
The attacker first constructs a \emph{topic-specific cache} by running the same model on a prompt reflecting the target topic.  
At a chosen trigger point, a contiguous block of the model’s active cache is overwritten with the corresponding block from the topic cache.  
By preserving structural alignment (identical tensor shapes and consistent time-step boundaries), the replacement remains coherent and allows the model to attend to this alternate internal history when generating subsequent tokens.


\begin{algorithm}[t]
\caption{History Swapping Attack}
\label{alg:history-swapping}
\begin{algorithmic}[1]
\Require Base model $M$, tokenizer, attacker topic prompt
\Require swap\_token, swap\_percent, layers\_affect\_percent, from\_beginning, max\_length
\State $\text{topic\_cache} \gets \text{RunModelAndRecordCache}(M, \text{attacker\_prompt})$
\State $\text{cache} \gets \text{InitializeEmptyCache}()$
\State $\text{output} \gets [\;]$
\For{$t = 1$ to max\_length}
    \State $\text{token} \gets M.\text{next\_token}(\text{cache})$
    \State Append($\text{output}, \text{token}$)
    \If{$\text{token} = \text{EOS\_token}$}
        \State \textbf{break}
    \EndIf
    \State $\text{cache} \gets \text{UpdateCache}(\text{cache}, \text{token})$
    \If{$t = \text{swap\_token}$}
        \State $T \gets \text{CurrentCacheLength}(\text{cache})$
        \State $m \gets \lfloor \text{swap\_percent} \cdot T \rfloor$
        \State $s \gets T - m$ \Comment{start index for swap}
        \State $\text{timesteps} \gets \{s, \ldots, T-1\}$ \Comment{last $m$ positions}
        \State $\text{topic\_timesteps} \gets \{0, \ldots, m-1\}$ \Comment{first $m$ from topic cache}
        \State $L \gets \text{TotalLayers}(M)$
        \State $k \gets \lfloor \text{layers\_affect\_percent} \cdot L \rfloor$
        \If{\text{from\_beginning}}
            \State $\text{layers} \gets \{0, \ldots, k-1\}$
        \Else
            \State $\text{layers} \gets \{L - k, \ldots, L-1\}$
        \EndIf
        \For{each $\ell \in \text{layers}$}
            \For{$i = 0$ to $m-1$}
                \State $\text{cache}[\ell][s + i] \gets \text{topic\_cache}[\ell][i]$
            \EndFor
        \EndFor
    \EndIf
\EndFor
\State \Return $\text{output}$
\end{algorithmic}
\end{algorithm}

Replacing a full, coherent block distinguishes History Swapping from noise-based corruption:  
instead of degrading the model’s internal state, the attack injects a structured alternative that can redirect topic, planning, and trajectory without weakening fluency.

\subsection{Control Parameters}
History Swapping presents several parameters for evaluation:
\begin{enumerate}
    \item \textbf{swap\_token} — the generation step at which the overwrite occurs.
    \item \textbf{swap\_percent} — the fraction of the most recent time steps replaced.
    \item \textbf{layers\_affect\_percent} — the proportion of transformer layers whose cache entries are modified.
    \item \textbf{from\_beginning} — whether the affected layers are the earlier or later in the network.
\end{enumerate}

These parameters allow for a systematic analysis of how different parts of the cache contribute to topic control, coherence, and long range planning during generation.

\section{Experimental Setup}
\begin{table}[h]
  \centering
  \caption{Evaluation parameters for History Swapping experiments}
  \label{tab:history-swapping-setup}
  \begin{tabular}{@{}lp{6cm}@{}} 
    \toprule
    Aspect & Details \\
    \midrule
    Models & Qwen3-32B, Qwen3-14B, \newline Qwen3-8B, Qwen3-4B-Instruct (2507) \\
    Attack Parameters & 
      \textbf{swap\_token:} 20\%, 40\%, 60\% \newline
      \textbf{swap\_percent:} 25\%, 50\%, 75\% \newline
      \textbf{layers\_affect\_percent:} 100\%, 25\% \newline
      \textbf{from\_beginning:} True/False \\
    Output Lengths & 1024, 2048, 4096 tokens \\
    Total Configurations & 324 \\
    \bottomrule
  \end{tabular}
\end{table}

As shown in Table~\ref{tab:history-swapping-setup}, we evaluate History Swapping across multiple models and attack parameter settings. 
All experiments are conducted using single-turn conversations with default decoding parameters. We use full BF16 checkpoints (no quantization) and operate in ``no-thinking'' mode without reasoning traces.
\newline
\\
\textbf{User prompt}:
\begin{quote}
    "Give a precise technical explanation of espresso extraction variables and expected effects."
\end{quote}
\vspace{1em}
\vspace{1em}
\textbf{Topic cache}:
\begin{quote}
    The topic\_cache is generated by running the model on:
    "Explain the lifecycle of a star, from nebula to supernova and the resulting black hole."
\end{quote}
To quantify topic drift, we constructed a focused corpus comprising star- and coffee-related texts. The text segments were embedded using the Qwen 3 8B embedding model, and topic centroids were calculated for each category.
A two-topic LDA model is trained on the corpus. During evaluation, a sliding window (60 tokens, stride 15) is applied to the output, with each window analyzed for cosine similarity to the star/coffee centroids and LDA topic scores.
Embeddings were projected onto the first two principal components, with LDA scores as a third axis, enabling 3D visualization of generation trajectories and systematic comparison of drift across attack configurations.

\section{Results and Discussion}
\subsection{Aggregate Deviations Across Models}
Out of the 324 history-swap configurations, 64 exhibited clear topic deviations, representing approximately 20\% of the runs. Table~\ref{tab:deviations} presents a summary of these deviations categorized by model size and target output length.

\begin{table*}[htbp]
\centering
\caption{Number of deviating configurations by model and target output length}
\label{tab:deviations}
\begin{tabular}{lrrrr}
\toprule
Model & 1024 tokens & 2048 tokens & 4096 tokens & Total \\
\midrule
Qwen 3 32B & 4 & 8 & 5 & 17 \\
Qwen 3 14B & 5 & 7 & 6 & 18 \\
Qwen 3 8B & 3 & 6 & 5 & 14 \\
Qwen 3 4B Inst & 3 & 7 & 5 & 15 \\
\midrule
Total & 15 & 28 & 21 & 64 \\
\bottomrule
\end{tabular}
\end{table*}

All deviating runs share a single condition: every transformer layer is overwritten (layers\_affect\_percent $=1.0$) for a given timestep.
Under these settings, full-layer attacks are the only ones that produce topic-level hijacks;
partial-layer attacks do not cause deviations in any configuration we evaluate.

\subsection{Qualitative Hijack Behaviors}
Among the 64 deviating runs, we observe three main behaviors:
\begin{enumerate}
    \item Immediate, persistent hijack: The topic shifts to the star lifecycle immediately following the swap and remains on the attack topic for the remainder of the sequence.
    \item Immediate hijack with partial recovery: The topic initially transitions to the star lifecycle but subsequently reverts to coffee, resulting in alternating content between the two topics.
    \item Delayed hijack: The model continues discussing espresso extraction for a substantial number of tokens post-swap before abruptly shifting toward the star lifecycle, indicating a delayed effect of the attack.
\end{enumerate}

\begin{figure}[h!]
  \centering
  \includegraphics[width=\linewidth]{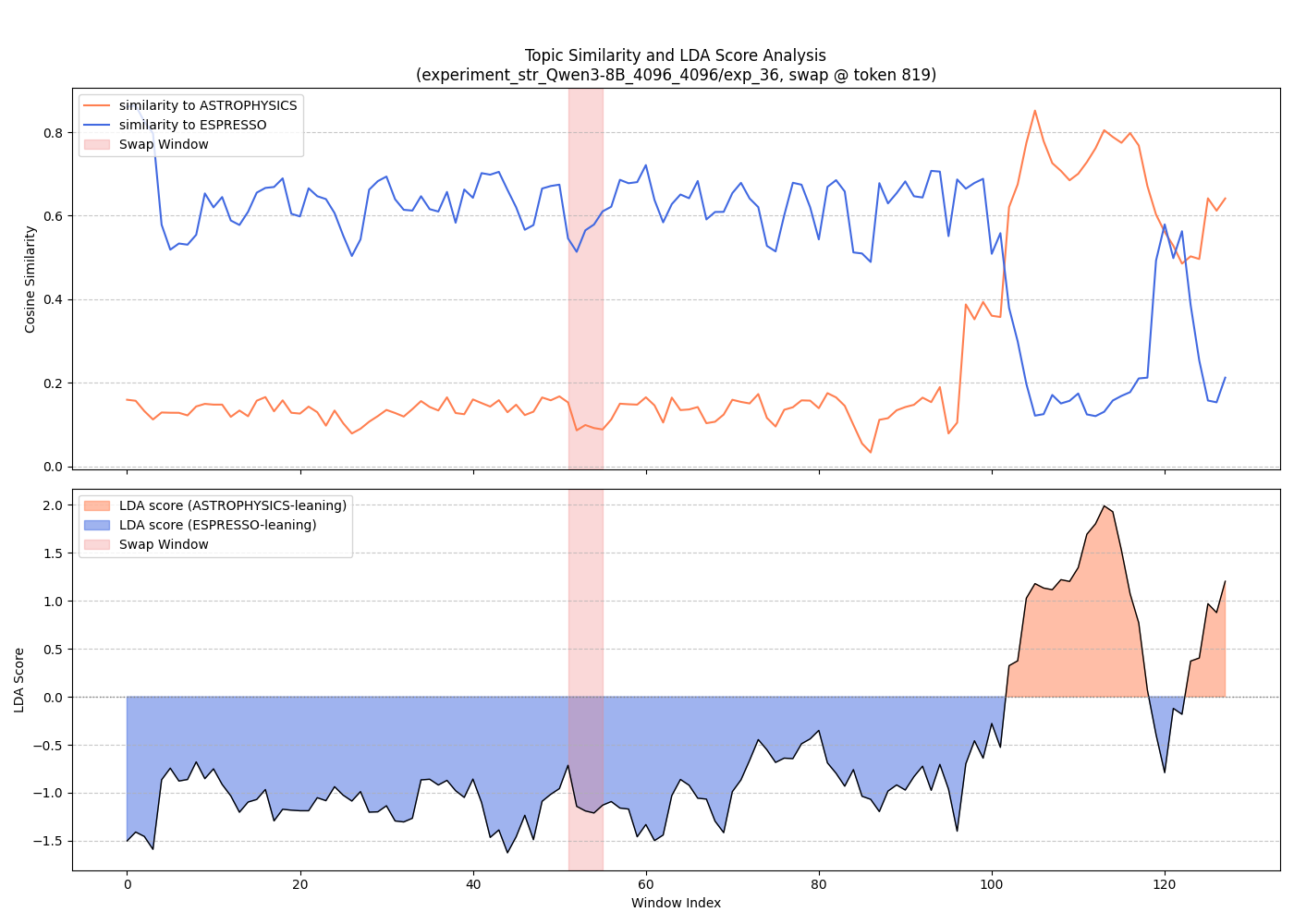}
  \caption{LDA Scores and Cosine Similarity for a delayed hijack attack}
\end{figure}

In some experiments, the topic remains unchanged, but the output degenerates into repetitive or incoherent text. In these cases, the altered cache disrupts the next-token distribution without producing a coherent shift toward the attack topic.

\subsection{Case Study: Qwen 3 8B, 2048-token Generation}

Using Qwen 3 8B with a target length of 2048 tokens, we examine how variations in swap\_percent alone can lead to markedly different outcomes when all other parameters are held constant. The controlled settings are

\begin{itemize}
    \item Model: Qwen 3 8B
    \item Target length: 2048 tokens
    \item swap\_token: 40\% of generation (approximately token 819)
    \item layers\_affect\_percent: 100\%
    \item from\_beginning: False
\end{itemize}
We vary only swap\_percent {25\%, 50\%, 75\%} and observe the outcomes summarized in Table~\ref{tab:swap_percent_effect}.
\begin{figure}[h!]
  \centering
  \includegraphics[width=\linewidth]{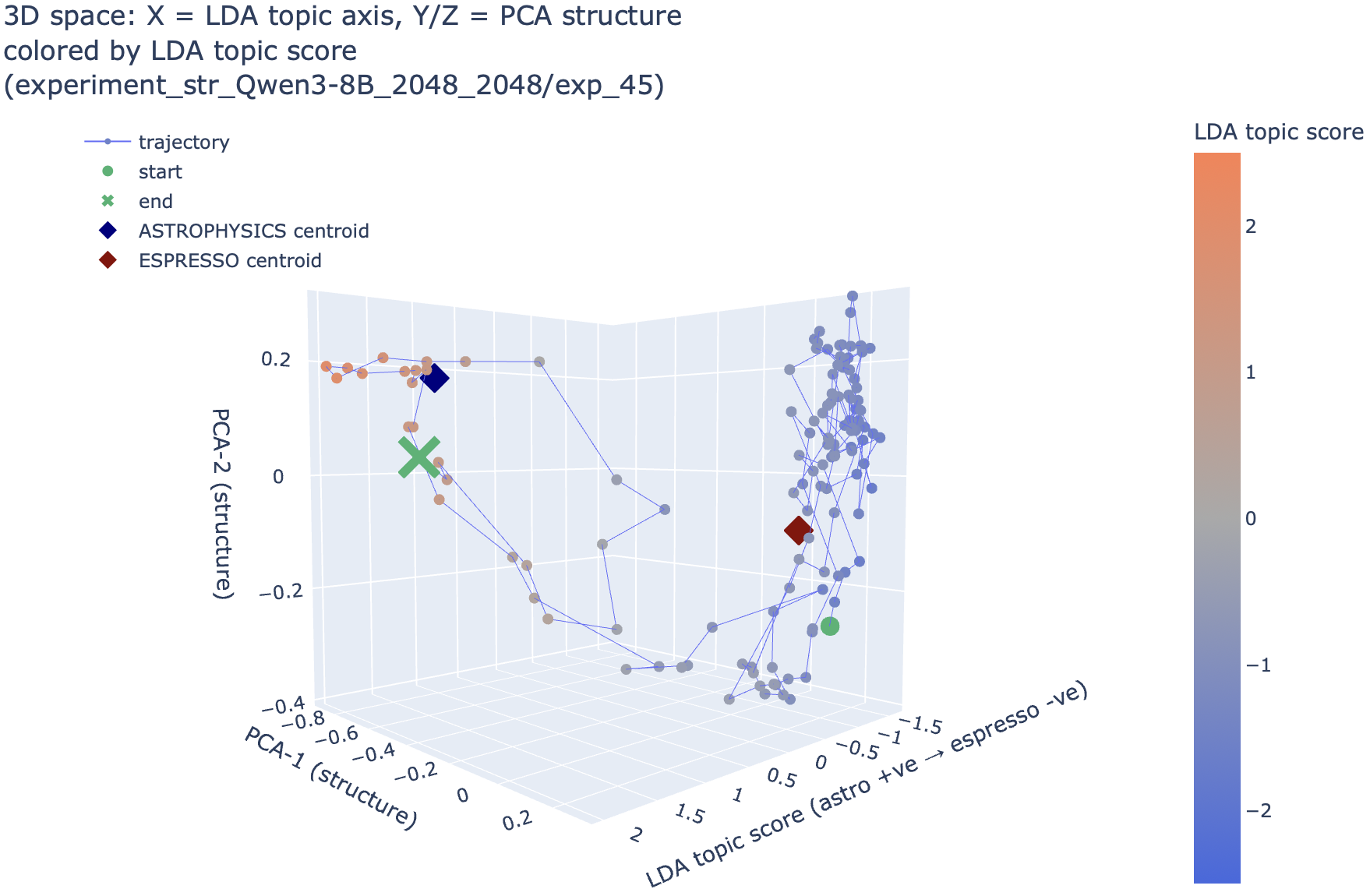}
  \caption{3D embedding trajectory plot for a delayed hijack attack(swap\_percent: 25\%)}
    \label{fig:8b_2048_2048_exp45}
\end{figure}

\begin{figure}[h!]
  \centering
  \includegraphics[width=\linewidth]{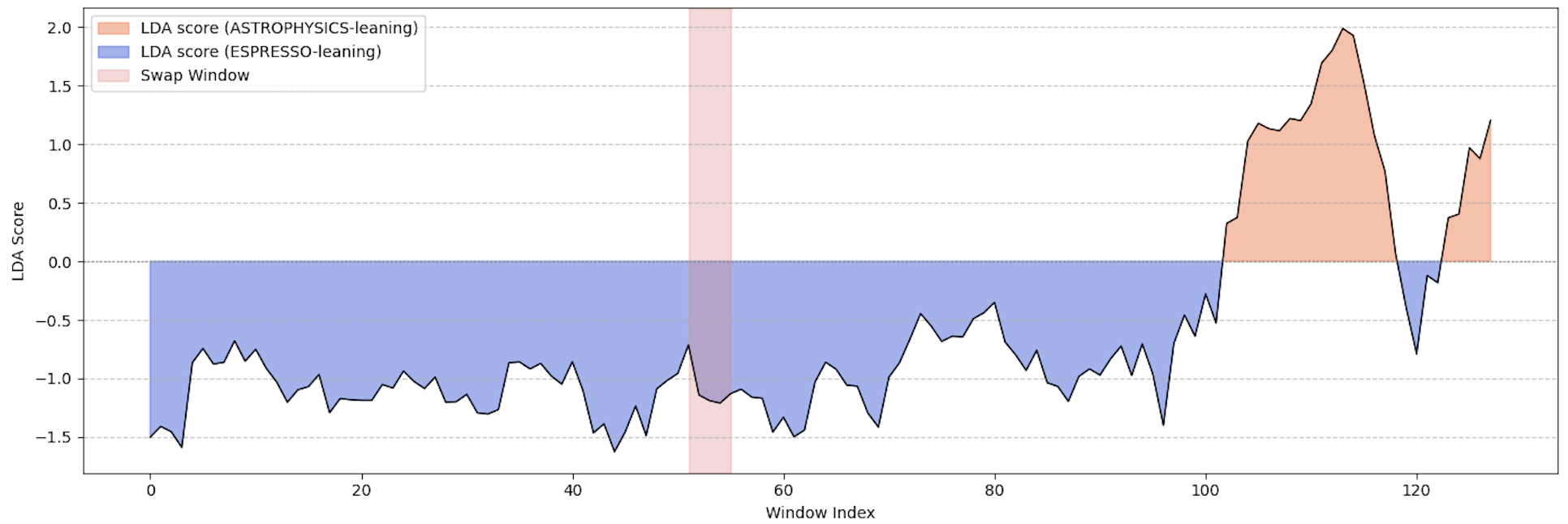}
  \caption{LDA Scores for a delayed hijack attack(swap\_percent: 25\%)}
    \label{fig:8b_2048_2048_exp45_lda}

\end{figure}

\begin{figure}[h!]
  \centering
  \includegraphics[width=\linewidth]{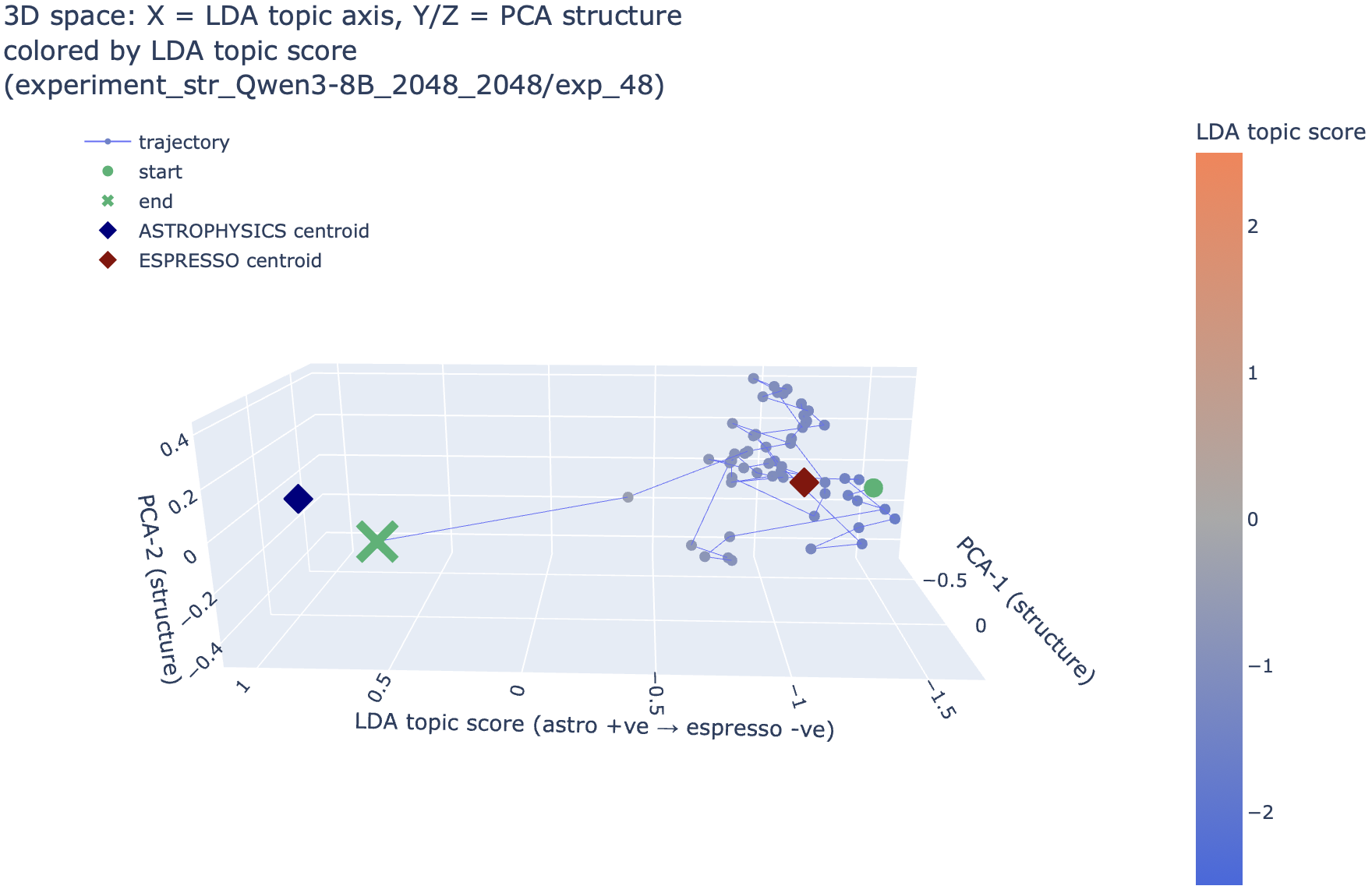}
  \caption{3D embedding trajectory plot depicting a model collapse scenario(swap\_percent: 50\%)}
      \label{fig:8b_2048_2048_exp48}

\end{figure}

\begin{figure}[h!]
  \centering
  \includegraphics[width=\linewidth]{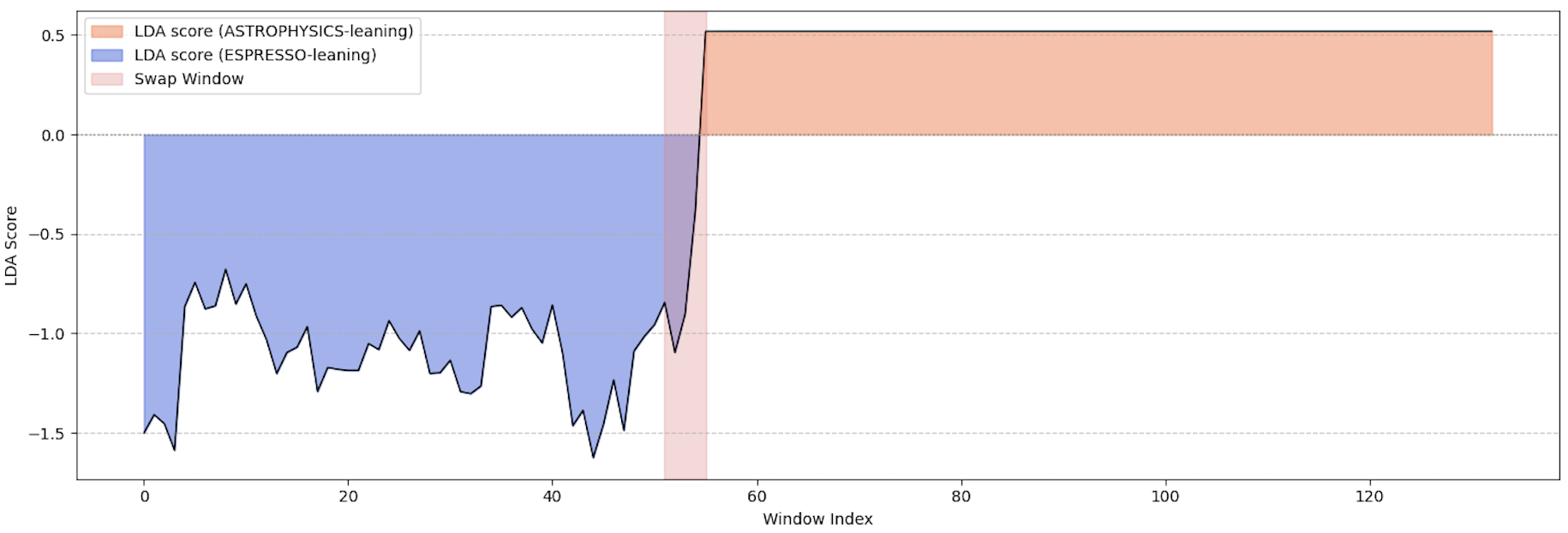}
  \caption{LDA Scores for a model collapse scenario(swap\_percent: 50\%)}
        \label{fig:8b_2048_2048_exp48_lda}

\end{figure}
\begin{figure}[h!]
  \centering
  \includegraphics[width=\linewidth]{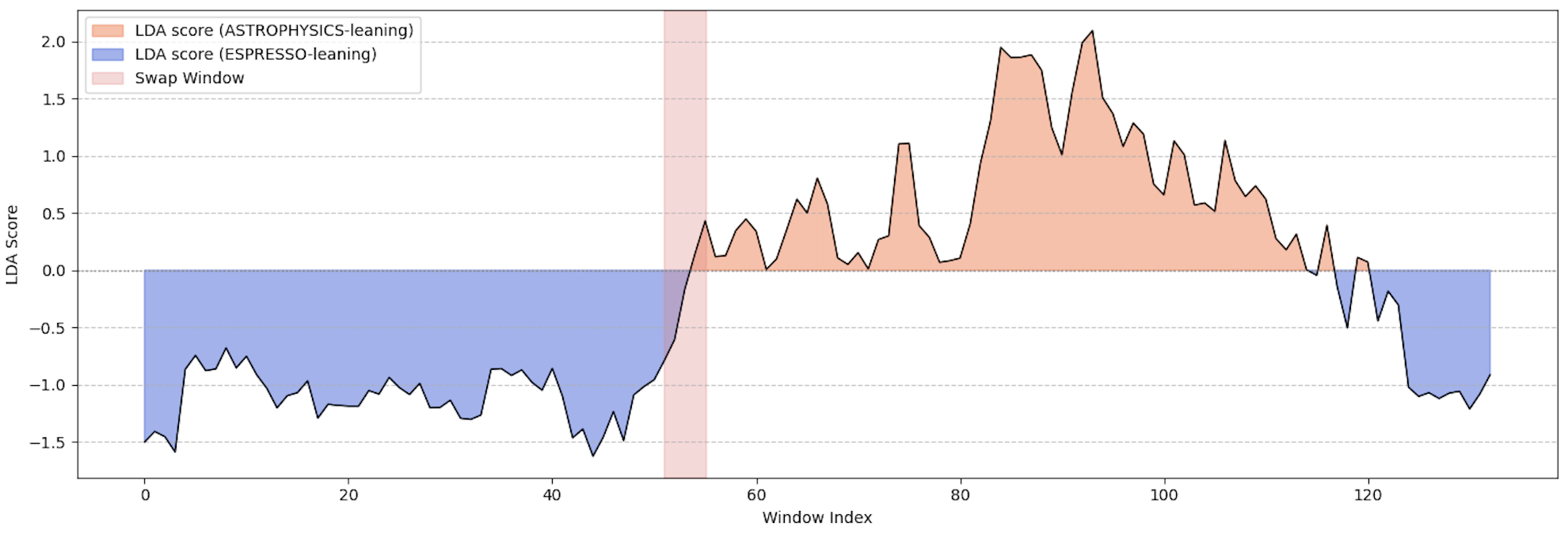}
  \caption{LDA Scores for an immediate hijack attack(swap\_percent: 75\%)}
          \label{fig:8b_2048_2048_exp51}

\end{figure}

\begin{figure}[h!]
  \centering
  \includegraphics[width=\linewidth]{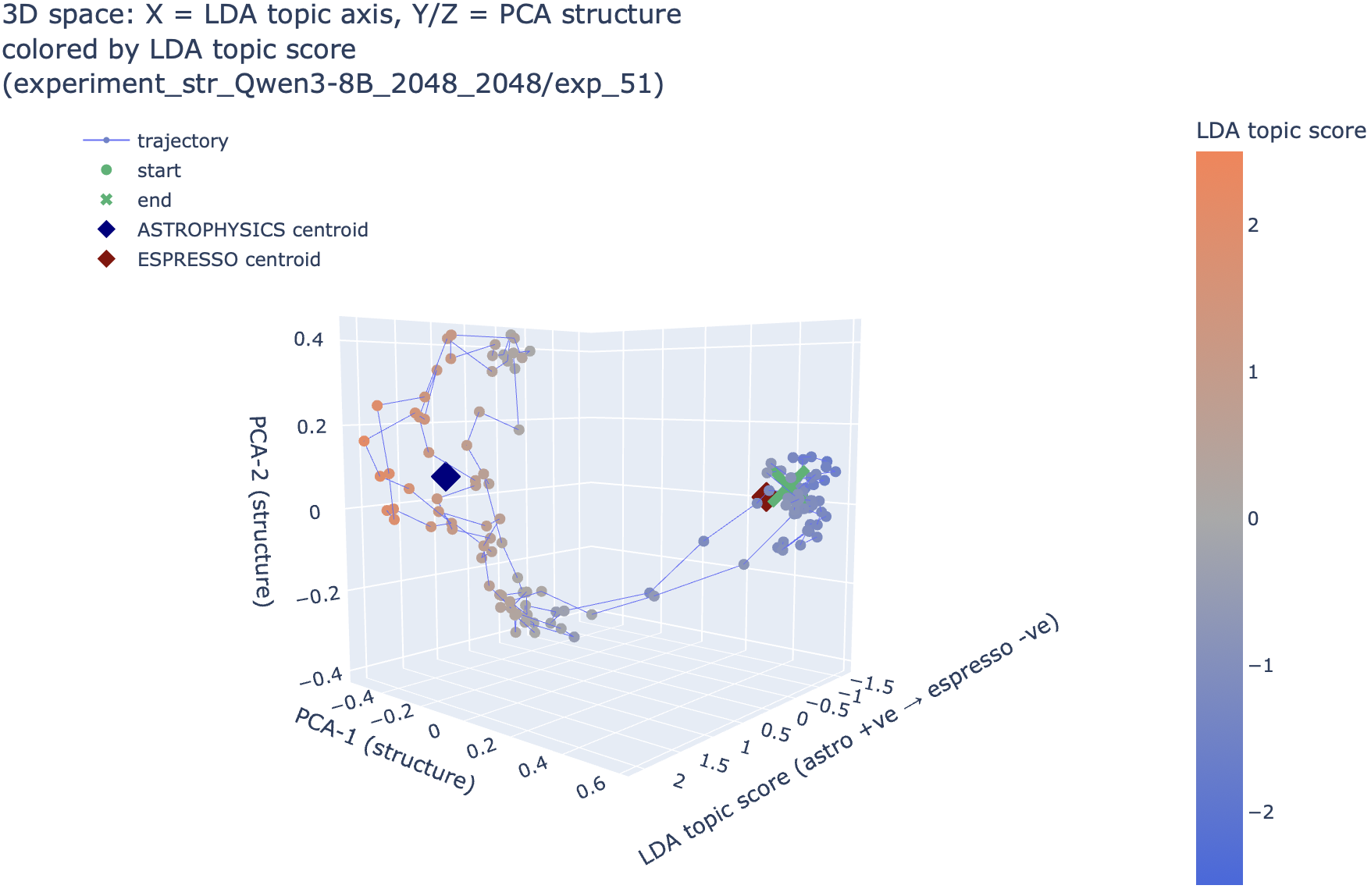}
  \caption{3D embedding trajectory plot for an immediate hijack attack(swap\_percent: 75\%)}
            \label{fig:8b_2048_2048_exp51_lda}

\end{figure}

\begin{figure}[h!]
  \centering
  \includegraphics[width=\linewidth]{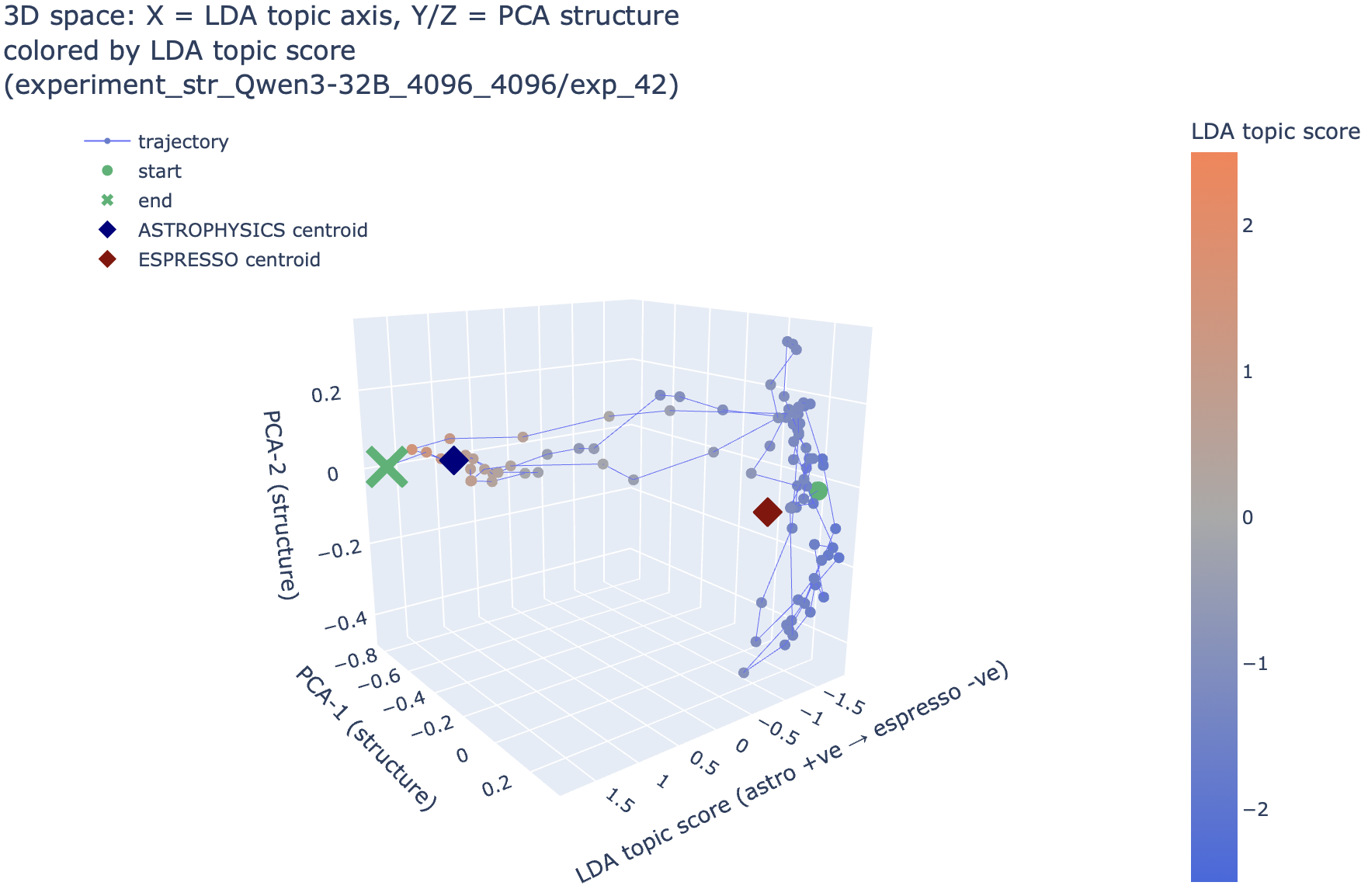}
  \caption{3D embedding trajectory plot, showing an alternating behavior}
              \label{fig:32b_4096_4096_exp42_lda}

\end{figure}

\begin{figure}[h!]
  \centering
  \includegraphics[width=\linewidth]{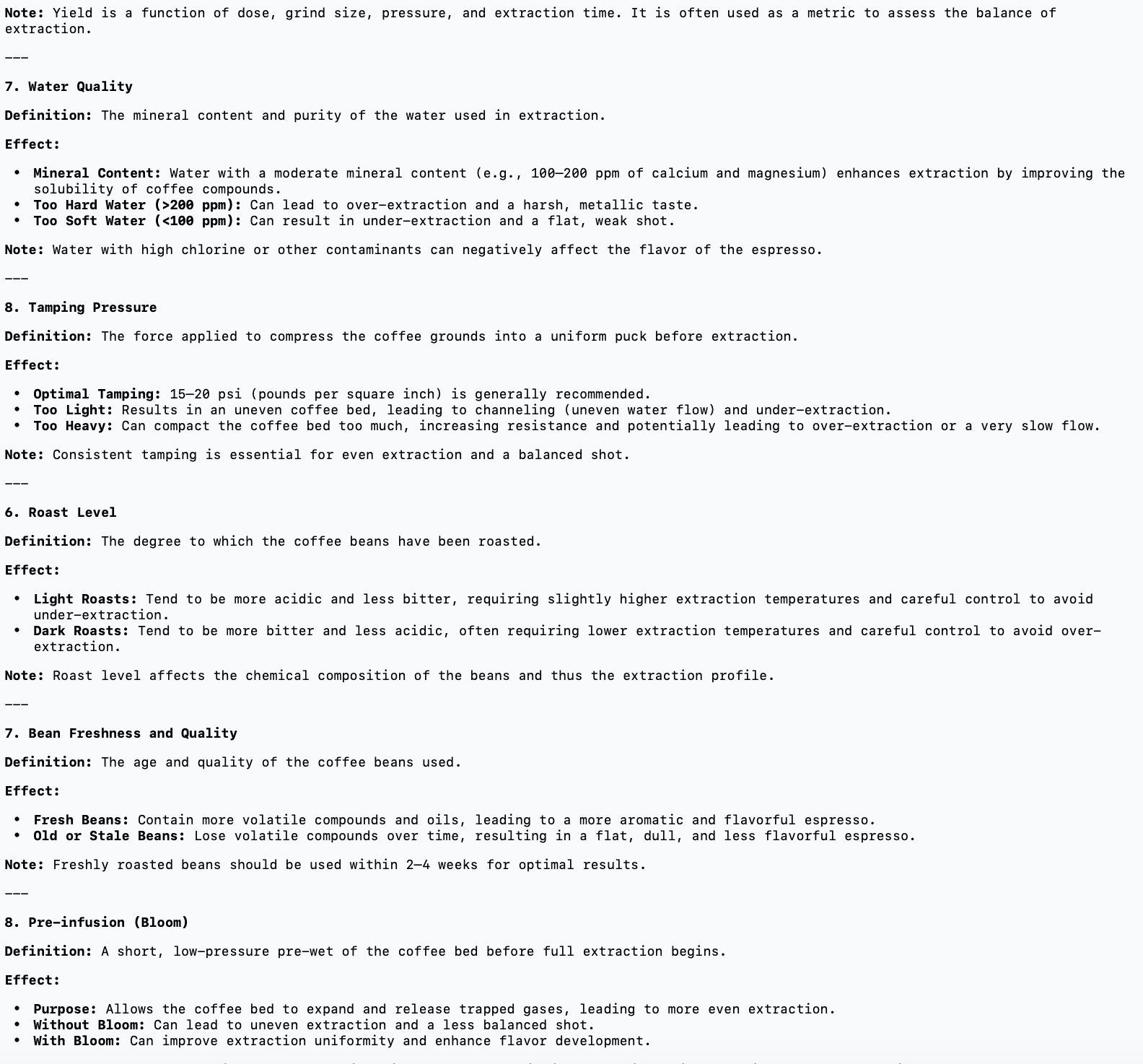}
  \caption{Text output from Qwen 3 14B showing duplicated list numbering after a history swap applied to later layers }
                \label{fig:erronous_text}

\end{figure}

\begin{table*}[htbp]
\centering
\caption{Effect of swap\_percent for Qwen 3 8B, 2048-token runs (swap\_token $\approx819$, 100\% layers, from\_beginning = False)}
\label{tab:swap_percent_effect}
\begin{tabular}{lp{0.7\linewidth}}
\toprule
swap\_percent & Observed behavior \\
\midrule
25\% & The model maintains a focus on coffee extraction for approximately 80 summarized in Table.
0 tokens after the swap before shifting to the star topic, with the conclusion incorporating references to both coffee and stellar evolution shown in figure ~\ref{fig:8b_2048_2048_exp45} and ~\ref{fig:8b_2048_2048_exp45_lda}.\\
\addlinespace
50\% & Generation quality collapses after the swap. The model repeats patterns (e.g., "1000...") until the token limit is reached shown in figure ~\ref{fig:8b_2048_2048_exp48} and ~\ref{fig:8b_2048_2048_exp48_lda}. \\
\addlinespace

75\% & The model transitions immediately to the star lifecycle, covering topics such as nebulae, supernovae, and black holes. Near the conclusion, it produces a structured summary table detailing coffee extraction variables and their effects shown in figure ~\ref{fig:8b_2048_2048_exp51} and ~\ref{fig:8b_2048_2048_exp51_lda}. \\

\bottomrule
\end{tabular}
\end{table*}

These runs illustrate how the magnitude of the overwrite influences behavior under fixed timing and depth: stable hijack, catastrophic collapse, or delayed topic shift.

\subsection{Persistence of Planned Structure}
Across both deviating and non-deviating experiment, the model consistently produces a structured summary table of espresso-extraction variables and effects toward the end of the sequence. This behavior persists even when swap\_percent reaches 75\% and the model generates extensive star-topic content after the swap; it can devote many tokens to the star lifecycle while still producing a coherent coffee-focused table, sometimes incorporating elements from both topics. Because the attack does not completely overwrite the entire history, the table structure and its association with the original prompt remain intact in the unmodified portions of the cache. This indicates that certain aspects of high-level response planning, such as the decision to include a summary table, are established early in generation and can survive relatively aggressive partial overwrites later in the sequence. This visualized in the 3D embedding trajectory plots shown in figure ~\ref{fig:8b_2048_2048_exp51_lda} and ~\ref{fig:32b_4096_4096_exp42_lda}.

\subsection{Layer Location and Numbering Behavior}
The from\_beginning parameter determines whether History Swapping affects early or late transformer layers, and this selection influences finer-grained structural behaviors even in the absence of observable topic deviations.
For Qwen 3 14B with a 2048-token target length, $swap\_percent = 75\%$, $layers\_affect\_percent = 25\%$, and $swap\_token \approx 1228$:

With $from\_beginning = False$ (later layers affected), the model outputs a numbered list (points 1-8).
After the swap at item 8, the list restarts with items 6, 7, and 8, causing duplicated numbering as shown in figure ~\ref{fig:erronous_text}.

With the same configuration and $from\_beginning = True$ (early layers affected), the model completes the list without disruption.

This pattern also occurs at $swap\_token \approx 819$ but not at $swap\_token \approx 409$. These patterns indicate that the final layer's KV states contribute to maintaining list order and local discourse structure, and that disruptions become more pronounced at higher token positions.

\subsection{Output Length in 4096-token Configurations}
\begin{figure}[h!]
  \centering
  \includegraphics[width=\linewidth]{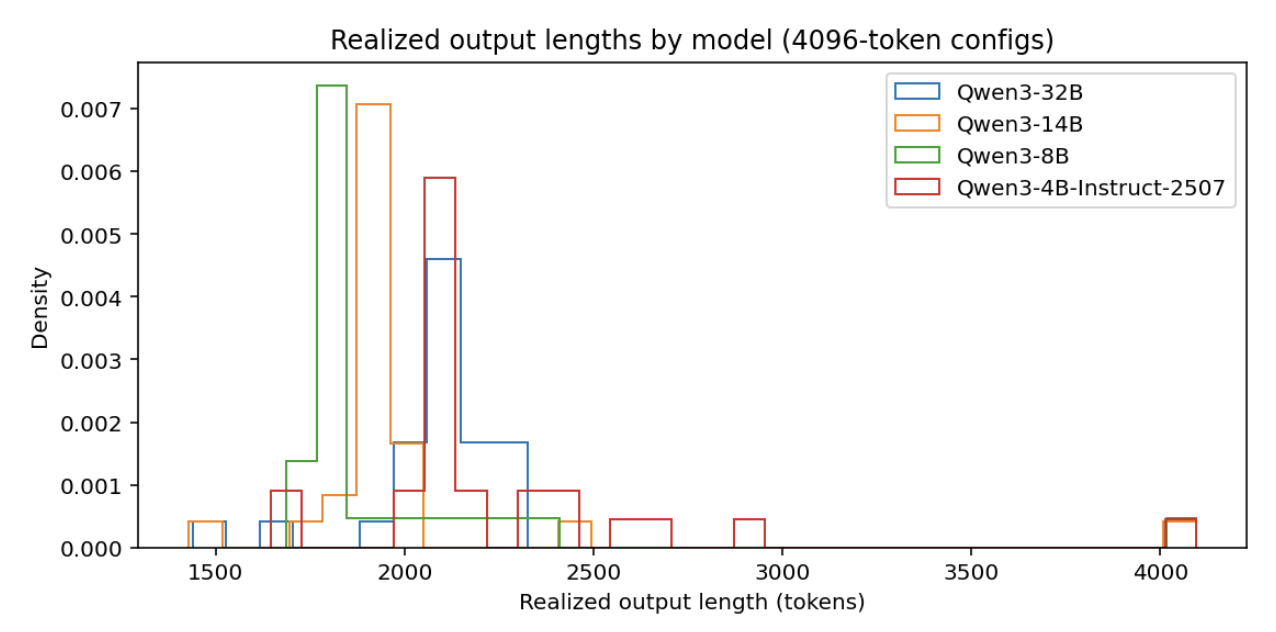}
  \caption{Histogram of realized output lengths for all 4096-token configurations}
  \label{fig:TokenPLot}
\end{figure}

Across the 108 experiments configured with a 4096 token maximum, actual outputs cluster around roughly 2000 tokens.
Realized lengths range from 1428 to 4096 tokens, with a mean of 2093 and a median of 2034 (25th percentile 1847, 75th percentile 2139).
This pattern is consistent across models, swap\_token settings, and both deviating and non deviating runs as shown in figure ~\ref{fig:TokenPLot}.

The observed consistency suggests that output length is primarily determined by early KV cache states, with mid-sequence modifications exerting limited influence on sequence termination. Factors such as the “no thinking” mode and default decoding parameters likely contribute to this behavior. We report this as an empirical observation specific to our experimental setup rather than a general property of Qwen 3.

\section{Limitations and Future Work}
This work constitutes an initial investigation into cache integrity attacks, with a focus on single-turn conversations and the behavior of a model within a single extended generation. A natural extension of this research is to evaluate History Swapping in multi-turn scenarios, particularly to determine whether “suppressed” attacks may reemerge in subsequent turns—analogous to delayed prompt injection but operating at the KV cache level. This consideration is increasingly important as deployed systems increasingly offload, reuse, and share KV state.
All experiments were conducted using full-precision BF16 checkpoints without quantization. In practical deployments, models frequently employ weight and activation quantization, and recent frameworks such as Unsloth\cite{unsloth} apply dynamic quantization to selected layers. Additionally, KV Cache Quantization is also a developing field trying to extend the context window of LLMs\cite{hooper2024kvquant}. Evaluating History Swapping across different quantization schemes would therefore enhance the generalizability of these findings.
Additionally, this study does not address KV cache compression or eviction mechanisms\cite{kang2024gear}, which could either mitigate or amplify the effects of History Swapping depending on how injected topic cache segments are treated.
Finally, our analysis is restricted to a single prompt pair (coffee versus stellar evolution), a Qwen 3 8B embedding coupled with a two-topic LDA pipeline, and a fixed family of block-level swaps at predetermined sequence fractions. Future work may broaden this framework by exploring alternative topics, classifiers, cache-editing strategies, model families, and targeted modifications to the “thinking” token space.

\section{Conclusion}
We introduced History Swapping as a KV cache integrity attack and evaluated it across 324 configurations on the Qwen 3 model family. In our experiments, only full-layer overwrites resulted in topic-level hijacks, whereas partial-layer modifications did not induce deviations. We identified three hijack behaviors—immediate, partial recovery, and delayed—as well as cases in which the model degenerated into low-quality repetitive output rather than exhibiting a topic shift.

Several observations indicate that structured information is encoded early in the generation. The persistence of summary-table generation despite substantial mid-sequence overwrites suggests that aspects of answer planning are established within the initial KV blocks. Experiments on numbering behavior further demonstrate that later-layer states play a role in preserving list order and discourse structure. Additionally, output lengths in 4096-token configurations cluster around approximately 2000 tokens, indicating that sequence termination is largely influenced by early cache states.
These findings demonstrate that the KV cache represents a significant internal state for both security analysis and mechanistic interpretability. In addition to prompts and model weights, the cache encodes planning structures and topic trajectories that can be selectively altered. Enhancing protections for KV cache access and deepening our understanding of its influence on model behavior will become increasingly critical as modern serving systems increasingly reuse and share cache states across sessions and hardware.

\bibliographystyle{ACM-Reference-Format}
\bibliography{sample-base}

\end{document}